**Effective management of white rust disease in red amaranth: a field study in Dhaka, Bangladesh**

**Short title: Management of white rust disease in red amaranth**

Abu Noman Faruq Ahmmed[1,] MD. Zahidul Islam[1], and Raihan Ferdous[1*]

[1]Department of Plant Pathology, Sher-e-Bangla Agricultural University, Dhaka-1207, Bangladesh

*Corresponding author: Raihan Ferdous

Email: raihanf.agri-2011106@sau.edu.bd

Phone: +880 1714 547678

**ABSTRACT**

This study aimed to evaluate the effective management strategies of *Albugo candida*, a pathogen of white rust disease in red amaranth (*Amaranthus tricolor* L.), accountable for the reduction of seed production. The study was performed during the Rabi season of 2018 and the Kharif season of 2019 at Sher-e-Bangla Agricultural University in Bangladesh. Eight treatments, including chemical, botanical, and biopesticide treatments such as Ridomil Gold 68 WG, Autostin 50 WP, Dithane M 45, Goldton 50 WP, the Bordeaux mixture, G-Derma, Garlic bulb extract, and Allamanda leaf extract, were evaluated. Four foliar sprays were applied at seven-day intervals after disease symptom onset. The field experiments followed a randomized complete block design with three replications. A microscopic study confirmed that *Albugo candida* was the causal organism. In both seasons, Ridomil Gold demonstrated superior efficacy in reducing disease incidence in plants, disease incidence in leaves, and disease severity, which were 63.07%, 62.78.5, and 84.31%, respectively, in Rabi and 69.73%, 65.71%, and 88.41%, respectively, in the Kharif season. Allamanda leaf extract also had statistically similar results, while Autostin exhibited promising effectiveness. Furthermore, compared with the other treatments, the combination of Ridomil Gold and Allamanda leaf extract significantly enhanced the growth parameters and seed yield in both seasons. Assessing the collective effectiveness of the treatments, Ridomil Gold demonstrated the most efficient control of white rust disease. Consequently, Ridomil Gold holds promise for application in red amaranth seed production. Additionally, the use of Allamanda leaf extract is an environmentally friendly approach to white rust disease management and promotes safer crop production practices. This study offers significant guidance to practitioners seeking optimal disease management strategies in red amaranth farming.

**Keywords:** *Albugo candida,* Allamanda leaf extract, *Amaranthus tricolor*, Bioagents, Botanicals, Disease management, Ridomil Gold.

# Introduction

Red amaranth (*Amaranthus tricolor L.*) is an important vegetable of the Amaranthaceae family. One of the widely favored leafy greens across South Asia, this plant, made its way to Africa, thriving near urban centers. Originating from India and spreading as far as China, it is also cultivated in the Caribbean and Pacific islands. With succulent, compact growth resembling spinach, it is a popular warm-season vegetable. Typically, cooked leaves are sometimes incorporated into salads, while the stems of this plant are eaten akin to those of asparagus in Indian cuisine [1]. It has been reported that vegetable crops need more intensive care and that they provide more profit than other crops, such as cereals, pulses, and oil seeds [2]. In Bangladesh, approximately 32000 acres of land are cultivated under red amaranth cultivation, which yields approximately 73000 m. ton. annually. Among the leafy vegetables grown in Bangladesh, red amaranth is the most common cultivated species, and its cultivation is increasing gradually [3]. Red amaranth is a good source of minerals such as calcium, magnesium, iron, zinc, manganese, and vitamins (A, B, and C). It offers a combined source of cheap and healthy food as well as medicinal value and health benefits [4].

Red amaranth cultivation technology is easy, but dozens of diseases, including fungal, bacterial, and viral diseases, cause significant yield loss under favorable conditions [5]. Among them, fungi are a great threat to red amaranth production and cause blight, dampening, white rust, stem rot, root knots, wilting, and other diseases. Among the diseases caused by red amaranth, white rust caused by *Albugo candida* is the major cause of rust because of its higher incidence, wider distribution, and adverse effects on the yield and quality of vegetable and seed production [6]. The oomycete fungus *A. candida* is known to be present wherever red amaranth is grown [7].

*A. candida* causes economically significant yield and quality losses in the seeds and vegetables of crucifers in several different ways. White rust disease on foliage reduces the photosynthetic capacity of plants and affects yield and normal plant development. Disease on foliage affects and degrades leaves for sale and human consumption [1].

The presence of conspicuous white pustules at the vegetative stage results in a blister-like appearance, and the leaves become twisted and deformed. It commonly occurs when nights are cool and damp and days are warm. The epidemic development of white rust is dependent upon different factors, viz. aggressiveness of races, amount of initial inoculum, time of first appearance of the disease, and prevailing weather conditions that determine the final intensity and severity of the disease [8].

The disease incidence and severity of white rust caused by red amaranth varied among the locations and seasons. Resistant cultivars, if available, are the most effective, practical, eco-friendly, and economical choice. However, at present, no such resistant varieties are available in Bangladesh [21]. Several chemicals such as Mersil, Manzate 200, Bayleton, Bromosan, Topsin M, and Kocide 101 have been evaluated for their ability to control this disease in India [9]. In Bangladesh, recommendations for the management of white rust are also limited. Very few studies have been conducted on managing this disease [22–23]. Several chemical fungicides, viz. Ridomil Gold 68 WP (Mancozeb+Metalaxyl) @ 0.2%, Contaf 25 EC (Hexaconazole) @ 0.1%, and X-tracare 300 EC (Propiconazole+Difenoconazole) 0.05% were reported to control white rust from red amaranth. Among these agents, Ridomil Gold was found to be the most effective at controlling white rust disease [10]. However, the antifungal effects of garlic extracts, allamanda extracts, and neem extracts have been reported by many researchers. Garlic bulb extract and Allamanda leaf extract were found to be promising for inhibiting the mycelial growth of fungi and reducing the disease incidence and severity in vegetables under field conditions [11]. In addition, *Trichoderma* species have been recognized as antagonists that can increase plant growth and induce plant resistance against fungal plant pathogens, such as *Albugo* spp. [12].

This disease is considered a major problem in the production of red amaranth seeds in Bangladesh. At the field level, farmers use chemical fungicides to manage this disease. However, the indiscriminate use of fungicides, frequent spring at high doses, and the use of two or more fungicides together are common practices for disease management in Bangladesh. Considering the economic importance of this crop, the present study aimed to identify the causal agent of white rust disease, along with its incidence and severity, and to evaluate the *in vivo* efficacy of fungicides, botanicals, and bioagents against white rust disease in red amaranth to ensure high-quality seed production.

## Materials and methods

### Study site and period

A field experiment was conducted at the Central Farm of Sher-e-Bangla Agricultural University (SAU), Dhaka-1207. A laboratory experiment was conducted at the Plant Disease Clinic of Sher-e-Bangla Agricultural University. The field experiment was conducted in the Rabi season (November-February) of 2018 and the Kharif season (March-June) of 2019. The study plot was a medium-high area belonging to the Modhupur tract under the Agro Ecological Zone (AEZ) 28. The soil texture was a silty loam, noncalcarious, dark gray soil from the Tejgaon soil series with a pH of 6.7.

### Test materials

The red amaranth variety used in the experiment was BARI Lalshak-1 (*Amaranthus tricolor*). This variety is open-pollinated and susceptible to white rust disease caused by red amaranth.

The seeds of red amaranth were collected from the Bangladesh Agricultural Development Corporation (BADC) office of Gabtoli, Dhaka, Bangladesh.

**Experimental design**

A randomized complete block design (RCBD) was used in this field study. The field was divided into 3 blocks and every block comprised nine unit plots thus the total number of plots was 27 (3×9=27). Tagging was applied to 9 different treatments, with a control in each block. The study was conducted with three replications. The treatments were distributed randomly in each block. The plot size was 3 m$^2$ (2 m × 1.5 m). The block-to-block distance was 75 cm, and the plot-to-plot distance was 50 cm.

**Soil preparation**

A tractor was used to cross-plow the soil twice to prepare it for seed germination. The weed was then taken out of the field. Following the Bangladesh Agricultural Research Council's 2018 guideline, fertilizer doses were applied during the final land preparation comprising 1 kg TSP, 1.2 kg MOP, 0.25 kg Gypsum, and half of 2 kg urea. During cultivation, the remaining urea fertilizer was applied in three separate applications. Additionally, 60 kg of cow dung was used in the field as an organic fertilizer to enhance the condition of the soil [26].

**Seed sowing**

After three days of land preparation, seeds were sown in the field. The plots of each block were selected randomly for the considered treatments. Seeds were sown in lines where line-to-line distance and plant-to-plant spacing were 15 cm and 12 cm respectively [27].

**Treatments**

A total of eight treatments were applied, including five chemical fungicides, two botanicals or plant extracts, and one bioagent, and were evaluated for their ability to treat white rust disease caused by red amaranth (Table 1). The treatments' application rates were followed, showing the best result against white rust disease demonstrated by different researchers [28–31]. In this research, the best fungicide was evaluated among the recommended doses of treatments.

**Field observation**

The experiment was conducted under natural epiphytic conditions. In 2018-19, Dhaka, Bangladesh experienced average weather conditions with a winter 2018 temperature of 20.5°C, 16.67 mm of precipitation, and 72% humidity, while summer 2019 saw a temperature of 30°C, 233.33 mm of precipitation, and 84.67% humidity [25]. Scouting and monitoring were performed regularly to determine disease symptoms on the leaves of the plants. The first disease incidence was observed 32 days after sowing (DAS) in the Rabi season and 28 DAS in the Kharif season.

**Spray schedule**

To control white rust disease, the abovementioned chemical fungicides, plant extracts, and bioagents were applied four times at 7-day intervals after disease onset. In the Rabi (winter) season of 2018, the first spray was applied 33 days after sowing (DAS), followed by 40, 47, and 54 DAS. In the Kharif (summer) season of 2019, the first spray was applied at 29 DAS, followed by 36, 43, and 50 DAS.

**Disease data recording and measurement**

From each plot, five plants were randomly selected for data collection. Disease incidence and severity data were recorded before the 1st spray and 7 days after the 1st, 2nd, 3rd, and 4th sprays for both seasons. A total of five parameters were considered to collect the data, viz. i.) Percent Disease Incidence in Plants (DIP), ii.) Percent Disease Incidence in Leaves (DIL), iii.) Percent disease severity (DS), iv.) Percent Disease Incidence Reduction over Control, and v.), Percent Disease Severity Reduction over Control [13–15].

The prevalence of white rust disease in red amaranth was recorded in terms of % disease incidence and % disease severity. After the disease symptoms appeared, disease incidence, disease severity, number of spots per leaf, and number of infected leaves per plant were recorded at regular intervals of seven days. Disease incidence and severity were calculated by the following formula [13-14]:

i. $\text{Disease incidence in plants (\%)} = \dfrac{\text{Number of infected plants}}{\text{Number of inspected plants}} \times 100$

ii. $\text{Disease incidence in leaves (\%)} = \dfrac{\text{Number of infected leaves}}{\text{Number of inspected leaves}} \times 100$

iii. $\text{Disease severity (\%)} = \dfrac{\text{Area of leaf tissues infected}}{\text{Area of leaf tissues inspected}} \times 100$

The reductions in disease incidence and severity relative to the control were calculated by using the following formula [15]:

iv. $\text{Disease incidence reduction over control (\%)} = \dfrac{C - T}{C} \times 100$

v. $\text{Disease severity reduction over control (\%)} = \dfrac{C - T}{C} \times 100$

Here, C = percent disease incidence or severity on the control plot, and T = percent disease incidence or severity on the treatment plot.

**Data recording and measurement of plant growth and yield**

The agronomic data on plant growth and yield were recorded to determine the effect of disease on seed production and yield contribution parameters of the crop. The data were collected from five tagged plants in each plot after harvesting. A total of seven agronomic parameters were considered for data collection, viz. i.) Total no. of plants per plot, ii.) Plant height (cm), iii.) number of leaves per plant, iv.) Number of inflorescences per plant, v.) Seed yield per plot (g), vi.) Seed yield per meter square area (g), and vii.) Seed yield increase compared with that of the control (%). The increase in seed yield compared with that of the control was calculated by using the following formula [15]:

x. Seed yield increase over control (%) = $\dfrac{T - C}{C} \times 100$

Here, T = yield of the respective treatment plot (g/m$^2$), and C = yield of the control plot (g/m$^2$).

**Identification of causal organisms**

The diseased leaves of the red amaranth plants were white pustules that were collected and observed under a compound microscope by preparing temporary slides. The slides were prepared from the diseased samples by picking, scraping, and sectioning methods to observe causal organisms under a compound microscope [24]. The pathogen was identified according to morphological studies based on the shape, color, and structural appearance of sporangia and mycelia of *A. candida*.

**Statistical analysis**

The collected data were statistically analyzed by the Statistics 10 computer package program. Analysis of variance (ANOVA) was used to determine the variation in the results from the study treatments. Treatment means were compared by the least significant difference (LSD) test.

**Results**

**Symptoms analysis**

Red amaranth plants grown in the experimental field at SAU under natural epiphytic conditions exhibited characteristic white rust symptoms (Figure 1) manifested as small, round to oval-shaped pustules (blister-like) on the lower leaf surfaces. Each pustule was approximately 1-3 mm in diameter and displayed a powdery white to creamy-colored outgrowth. As the disease progresses, multiple pustules coalesce, forming larger, irregularly shaped structures. Initially, only the lower leaves were affected. However, with time, the disease spreads to younger leaves, buds, and inflorescences. As the infection intensified, discoloration appeared on the upper leaf surface, corresponding to the white blisters below. This discoloured area lost photosynthetic capacity due to chlorosis, ultimately leading to leaf death. Severely infected leaves showed necrosis and became desiccated and detached from the plants.

**Identification of the causal organism**

*Albugo candida* was identified by microscopic examination as the causative agent of white rust disease in red amaranth. The mycelial growth of the infected leaves was observed under a compound microscope, revealing numerous sporangia and chains of sporangia associated with *A. candida*. These sporangia were hyaline and had a globose to spherical shape. Additionally, the endophytic mycelia appeared to be branched and hyaline (Figure 2).

**Disease management in the Rabi season, 2018**

**Measurement of disease incidence and severity during the Rabi season**

The effectiveness of various treatments against white rust disease in red amaranth plants was evaluated under natural conditions in the experimental field of Sher-e-Bangla Agricultural University in the Rabi season of 2018. Disease incidence and severity were assessed before the initial spray application (treatments) and at seven-day intervals following each subsequent spray. At 33 DAS (before the first spray), the disease incidence in plants (DIP), disease incidence in leaves (DIL), and disease severity (DS) were measured for each treatment plot. These values ranged from 7% to 10%, 6.33% to 10%, and 0.05% to 0.12%, respectively (Table 2). By 40 DAS (before the second spray), the impacts of the different treatments on DIP and DS varied significantly, ranging from 13.0% to 24.0% and 0.93% to 1.83%, respectively. Ridomil Gold (T1) demonstrated superior effectiveness compared to the other treatments in terms of these parameters. However, DIL measurements (13.33% to 25.67%) did not significantly differ among the treatments (Table 3). At 47 DAS (before the third spray), Ridomil Gold (19% DIP, 20% DIL, and 1.4% DS) exhibited significantly better results than did all the other treatments, except for Allamanda leaf extract, which displayed statistically similar results only for DS (1.6%) (Table 4).

The ability of the various treatments to control white rust disease differed significantly at 54 DAS (before the fourth spray). In terms of DIP and DIL, Ridomil Gold (26% and 22.23%, respectively) yielded the most favorable outcome. Additionally, the lowest DS was observed with Ridomil Gold (1.76%), which was statistically similar to that of Allamanda leaf extract (2.16%), followed by that of Autostin (3.33%) (Table 5). At 61 DAS (after 7 days of the fourth spray), only the Ridomil Gold treatment (24%) displayed significantly better results in terms of DIP among all the treatments. Similarly, DIL also varied significantly between treatments, with the lowest DIL recorded in Ridomil Gold (22.33%), which was statistically similar to that in the Allamanda leaf extract (28.33%). Moreover, the lowest DS was again observed for Ridomil Gold (1.83%), which was significantly similar to that of Allamanda leaf extract (2.43%), followed by that of the Bordeaux mixture (4.5%) (Table 6). Unsurprisingly, the control plot consistently exhibited the highest disease incidence and severity throughout the evaluation period (Tables 3-6).

**Reduction in disease incidence and severity during the Rabi season**

After seven days of final (fourth) spray, compared with the control treatment, the Ridomil Gold treatment had the greatest reduction in DIP (63.07%), DIL (62.78%), and DS (84.31%), followed by the Allamanda leaf extract and Autostin treatment (Table 7).

**Effects of different treatments on plant growth and yield in the Rabi season**

Data on the yield-contributing growth characteristics and seed yield of red amaranth were collected from each plot during the harvesting period of the Rabi season in 2018. Among all the treatments, Ridomil Gold had the best results for all the parameters of growth and yield characteristics, which were the number of plants per plot 122.67, plant height 64.66 cm, number of leaves per plant 10.33, number of inflorescences per plant 8.33, seed yield per plot 253 g, seed yield per $m^2$ area 84.33 g and seed yield increase over the control 53.32% (Table 8).

**Disease management in Kharif Season, 2019**

**Measurement of disease incidence and severity in the Kharif season**

Like in the Rabi season, disease incidence and severity were measured under each treatment plot at 29 DAS (before the first spray) and ranged from 5% to 7%, 5.66% to 9%, and 0.07% to 0.13%, respectively (Table 2). By 36 DAS (before the second spray), the lowest DIP was observed with Dithane M 45 (8.33%) and Allamanda leaf extract (8.33%) application, while DIL did not significantly differ among the treatments. However, the lowest statistically significant difference was detected for DS, which was achieved with Ridomil Gold (0.93%) (Table 3). During the period 43 DAS (before the third spray), the lowest disease incidence was observed for Ridomil Gold (15% DIP and 24% DIL) and Allamanda leaf extract (14.67% DIP and 23.33% DIL), with similar results shown for Dithane M 45 (28.33%) only for DIL. However, only Allamanda leaf extract (1.5%) had the best effect on the DS (Table 4). At 50 DAS (before the fourth spray), the Ridomil Gold (20% DIP, 27% DIL, and 2.16% DS) and Allamanda leaf extract (20% DIP, 25.66% DIL, and 2% DS) treatments produced significantly better results than did the other treatments (Table 5). At 57 DAS (after seven days of the fourth spray), Ridomil Gold (18.66% DIP, 24% DIL, and 1.93% DS) and Allamanda leaf extract (22% DIP, 27.66% DIL, and 2.23% DS) were the most effective disease control agents (Table 6), while the control plot was the least effective throughout the evaluation period (Tables 3-6).

**Reduction in disease incidence and severity in the Kharif season**

After seven days of final spray treatment, the greatest reductions in DIP, DIL, and DS were detected in the Ridomil Gold treatment group (69.73%, 65.71%, and 88.41%, respectively), followed by the Allamanda leaf extract and the Autostin spray plot (Table 7). Considering all the findings, the Ridomil Gold-treated plot showed the greatest reduction in disease incidence compared to that of the control.

**Effects of different treatments on plant growth and yield in the Kharif season**

During the harvesting period of the Kharif season, in 2019, among all the treatments, Ridomil Gold again had the best results for all the parameters of growth and yield characteristics, which were the number of plants per plot (116), plant height (59.16 cm), number of leaves per plant (9.33), number of inflorescences per plant (10), seed yield per plot (232 g), seed yield per m2 area (77.33 g), and increase in yield compared with the control (51.63%) (Table 9). Considering the overall growth and yield performance, the Ridomil Gold plot exhibited the best results, followed by the Allamanda leaf extract-sprayed plot.

# Discussion

During the course of the field investigation, it was noted that infected leaves exhibited rounded to oval-shaped, creamy white, small blister-like pustules predominantly on their lower surfaces, occasionally extending to the upper surfaces as the disease advanced. In severe instances, infected leaves and inflorescences desiccated and detached from the plants. Analogous findings were reported by Verma in 2012 in his studies on white rust disease in various cruciferous crops, indicating the occurrence of white to creamy-colored pustules primarily on the underside of leaves [16], while in 2009, Mishra *et al.* documented systemic infections manifesting in necrotic and desiccated inflorescences and terminal leaves [17].

Microscopic examination revealed that the sporangia of *Albugo candida* exhibited a hyaline and globose to spherical morphology, while the mycelia appeared endophytic, branched, and hyaline, findings consistent with those documented by Mirzaee *et al.* in 2009 [18].

In each of the two seasons (Rabi 2018 and Kharif 2019), four sprays were applied, comprising eight selected treatments (excluding the control), at seven-day intervals. The data were collected prior to the initial application and at seven-day intervals thereafter. Analysis of the collected data indicated that among the treatments, Ridomil Gold had the lowest incidence and severity of disease, closely followed by Allamanda leaf extract. Compared with alternative treatments, both treatments exhibited superior efficacy in reducing disease incidence and severity. Furthermore, in terms of their impact on plant growth characteristics and seed yield performance, plots treated with Ridomil Gold exhibited the most favorable outcomes, followed by those treated with Allamanda leaf extract. These findings are consistent with those reported by other researchers in the same field who identified Ridomil Gold as the most efficacious chemical fungicide for controlling white rust disease in red amaranth [10]. Similarly, in 2009, Huq and Rahman demonstrated a reduced disease incidence in plots treated with Ridomil Gold under natural conditions, followed by Sunvit [19]. In 2004, Mostafa reported the efficacy of Allamanda leaf extract in reducing disease incidence and severity in mustard crops, accompanied by enhanced plant growth parameters [20].

Considering the comprehensive performance evaluation of the treatments, Ridomil Gold emerged as the most effective chemical for controlling white rust disease caused by *Albugo candida*. Consequently, Ridomil Gold could be recommended for use in red amaranth seed production. Additionally, Allamanda leaf extract is an environmentally friendly approach for white rust disease management and is conducive to safe crop production practices.

## Conclusion

In conclusion, this study elucidated the disease symptomatology, identified the causal organism, and evaluated treatments for white rust disease caused by red amaranth. These findings underscore the efficacy of the Ridomil Gold and Allamanda leaf extracts in mitigating disease incidence and severity, as well as promoting favorable growth parameters and seed yield. The study recommends the adoption of Ridomil Gold as a chemical fungicide, while Allamanda leaf extract is suggested as an eco-friendly alternative in disease management practices for red amaranth cultivation. Moving forward, continued exploration of integrated pest management strategies is crucial for bolstering the resilience of red amaranth production systems against white rust disease.

**Author contribution statement**

**ANF Ahmmed** = Supervision, methodology, investigation; conceptualization, funding acquisition, validation, writing – review and editing; **MZ Islam** = Conceptualization, data curation, formal analysis, investigation, methodology, resources, software, validation, visualization, writing– original draft; **R Ferdous** = Data curation, formal analysis, resources, software, validation, visualization; writing– original draft, writing – review and editing. All the authors reviewed the manuscript and provided their feedback.

**Acknowledgments**

The author is grateful to the Ministry of Science and Technology, Government of People's Republic of Bangladesh, for providing the National Science and Technology (NST) Fellowship to the researcher for this research work in the year 2018-19. The author is also thankful to the Bangladesh Agricultural Development Corporation (BADC) for supplying the seeds.

**Funding information**

No external funding was received to support this work.

**Compliance with ethical standards**

**Conflict of interest**

The authors declare that there are no conflicts of interest.

**Ethical approval**

This study does not include human or animal subjects.

31. Cumagun CJR. Managing Plant Diseases and Promoting Sustainability and Productivity with Trichoderma: The Philippine Experience. J. Agr. Sci. Tech. 2012;14: 699–714. Available from: https://jast.modares.ac.ir/article-23-1525-en.pdf19

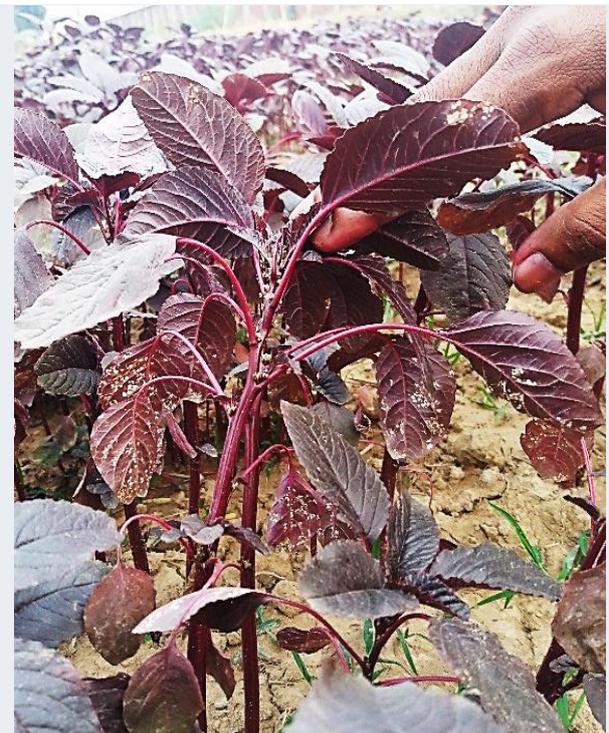
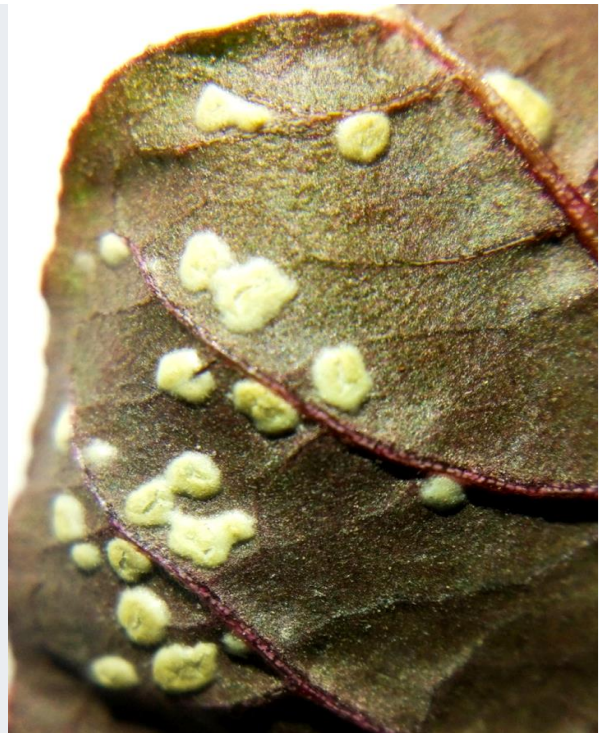

**Symptoms in standing plants**  **White to creamy colored powdery outgrowth**

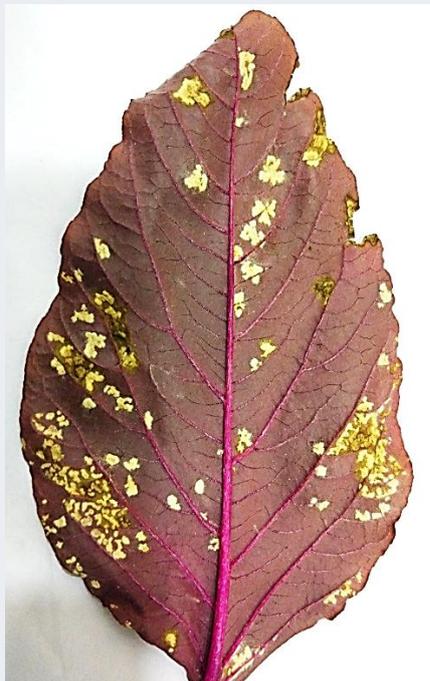
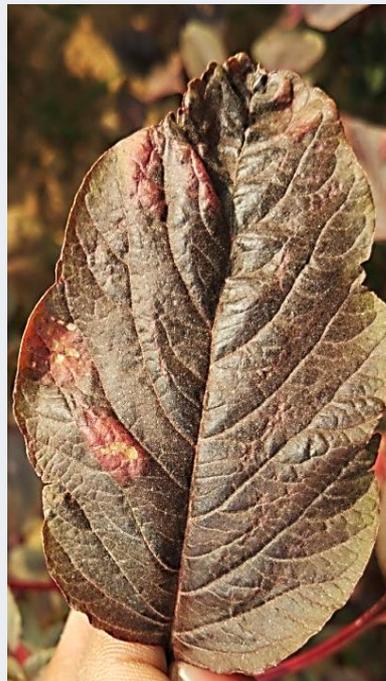
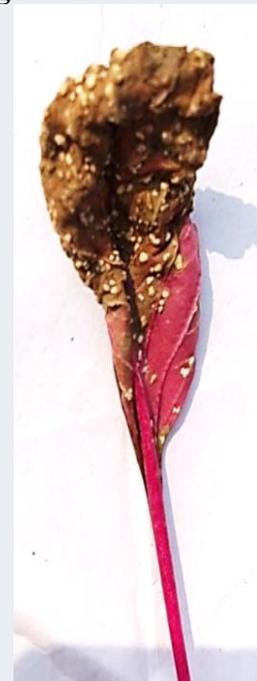

**Blister-like pustules on the lower surface of the leaf**  **Chlorosis on the upper leaf surface**  **Necrosis on severely infected leaf**

**Figure 1.** Symptoms of white rust disease in red amaranth plants were observed under field conditions, and the plants on both sides of the leaves exhibited typical characteristics.

| 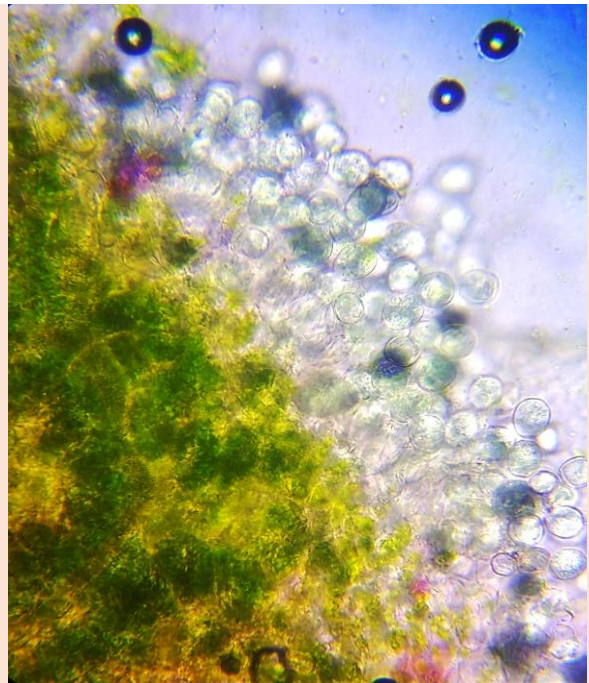 | 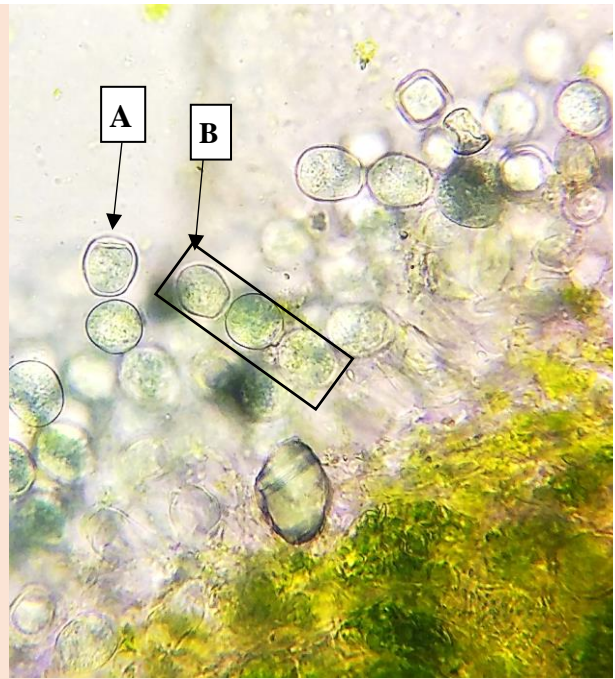 |
|---|---|
| **Section of infected host tissues** | **A. Loose sporangia; B. Chain of sporangia** |
| 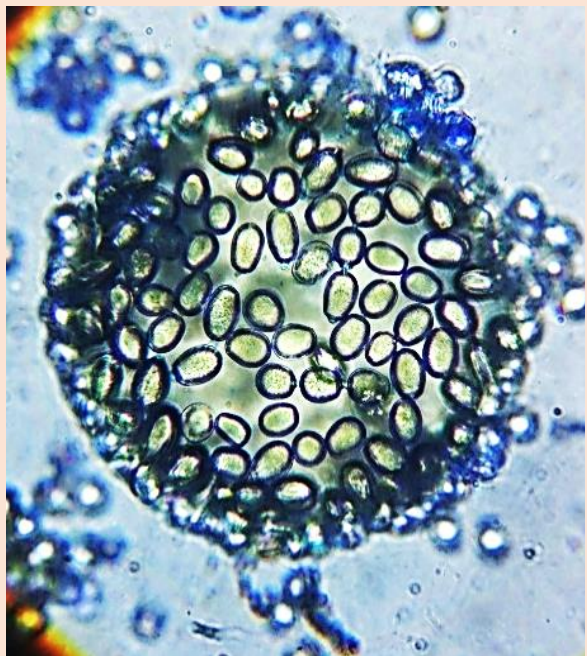 | 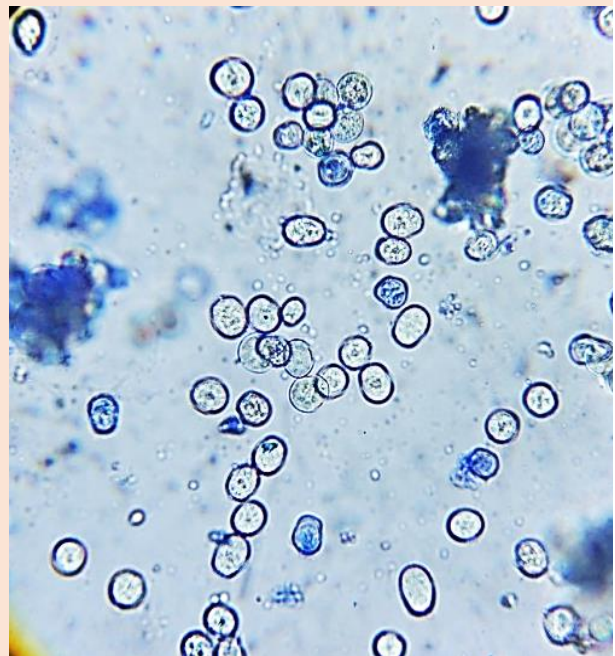 |
| **Globose sporangia** | **Asexual spore** |

**Figure 2.** Sporangia (asexual spore) of *Albugo candida* observed under a compound microscope (40X).

**Table 1.** Details of the treatments used for management of white rust disease of red amaranth

| Sl. no. | Trade name/ name | Common name / active ingredient | Rate of application (%) |
|---|---|---|---|
| 1. | **Ridomil Gold 68 WG** | Mencozeb+ Metalaxyl | 0.2 |
| 2. | **Autostin 50 WP** | Carbendazim | 0.2 |
| 3. | **Dithane M 45** | Mancozeb | 0.2 |
| 4. | **Goldton 50WP** | Copper oxychloride | 0.2 |
| 5. | **Bordeaux Mixture** | Copper sulfate and calcium hydroxide | 1 |
| 6. | **G-derma** | Bio agent (*Trichoderma* sp.) | 2 |
| 7. | **Garlic bulb extract** | *Allium sativum*; Botanicals | 3 |
| 8. | **Alamanda leaf extract** | *Allamanda cathartica* Botanicals/ plant extract | 3 |

**Table 2. Disease incidence and severity of white rust of red amaranth before treatments (1st spray) in the Rabi season of 2018 and the Kharif season of 2019.**

| Treatments | Rabi season, 2018 | | | Kharif season, 2019 | | |
|---|---|---|---|---|---|---|
| | Disease incidence in | | Disease severity (%) | Disease incidence in | | Disease severity (%) |
| | Plant (%) | Leaf (%) | | Plant (%) | Leaf (%) | |
| T$_1$ (Ridomil Gold) | 8.33 ab | 10.00 a | 0.11 a | 5.00 | 8.00 a-c | 0.13 |
| T$_2$ (Autostin 50WP) | 8.33 ab | 6.33 b | 0.09 a | 7.00 | 5.66 d | 0.07 |
| T$_3$ (Dithane M 45) | 10.00 a | 6.67 b | 0.10 a | 5.66 | 7.00 b-d | 0.11 |
| T$_4$ (Goldton 50WP) | 8.33 ab | 6.33 b | 0.10 a | 6.00 | 8.33 ab | 0.10 |
| T$_5$ (Bordeaux Mixture) | 8.67 ab | 8.33 ab | 0.11 a | 6.66 | 8.00 a-c | 0.10 |
| T$_6$ (G-derma) | 7.00 b | 7.00 b | 0.09 a | 5.66 | 6.66 a-c | 0.10 |
| T$_7$ (Garlic bulb extract) | 7.67 ab | 7.00 b | 0.05 b | 5.00 | 7.66 a-c | 0.12 |
| T$_8$ (Allamanda leaf extract) | 8.67 ab | 8.33 ab | 0.12 a | 6.33 | 9.00 a | 0.13 |
| T$_9$ (Control) | 10.00 a | 10.00 a | 0.10 a | 6.33 | 7.33 bc | 0.10 |
| LSD (0.05) | 2.70 | 2.63 | 0.03 | 2.32 | 1.39 | 0.04 |
| CV (%) | 18.27 | 19.58 | 19.81 | 22.51 | 10.71 | 23.43 |

Means followed by the same letters in a column did not differ at 5% level of significance by LSD.

**Table 3. Disease incidence and severity of white rust of red amaranth before 2nd spray in the Rabi season of 2018 and the Kharif season of 2019.**

| Treatments | Rabi season, 2018 | | | Kharif season, 2019 | | |
|---|---|---|---|---|---|---|
| | Disease incidence in | | Disease severity (%) | Disease incidence in | | Disease severity (%) |
| | Plant (%) | Leaf (%) | | Plant (%) | Leaf (%) | |
| $T_1$ (Ridomil Gold) | 13.00 c | 13.33 b | 0.93 d | 10.00 bc | 15.00 ab | 0.93 d |
| $T_2$ (Autostin 50WP) | 16.67 bc | 15.67 b | 1.16 cd | 11.33 ab | 14.33 b | 1.50 b-d |
| $T_3$ (Dithane M 45) | 18.67 b | 14.6 b | 1.50 a-c | 8.33 c | 17.66 ab | 1.66 bc |
| $T_4$ (Goldton 50WP) | 16.00 bc | 13.33 b | 1.67 ab | 11.00 ab | 13.33 b | 1.83 b |
| $T_5$ (Bordeaux Mixture) | 20.00 ab | 16.00 b | 1.33 b-d | 12.66 a | 15.33 ab | 1.83 b |
| $T_6$ (G-derma) | 18.67 b | 17.33 b | 1.33 b-d | 10.00 bc | 20.00 ab | 2.16 b |
| $T_7$ (Garlic bulb extract) | 19.00 ab | 15.67 b | 1.16 cd | 12.00 ab | 18.33 ab | 1.83 b |
| $T_8$ (Allamanda leaf extract) | 17.00 bc | 14.33 b | 1.00 d | 8.33 c | 18.33 ab | 1.00 cd |
| $T_9$ (Control) | 24.00 a | 25.67 a | 1.83 a | 13.33 a | 22.66 a | 3.33 a |
| LSD (0.05) | 5.11 | 6.17 | 0.48 | 2.36 | 5.67 | 0.66 |
| CV (%) | 16.32 | 22.00 | 21.23 | 12.66 | 19.04 | 21.59 |

Means followed by the same letters in a column did not differ at 5% level of significance by LSD.

Table 4. Disease incidence and severity of white rust of red amaranth before 3$^{rd}$ spray in the Rabi season of 2018 and the Kharif season of 2019.

| Treatments | Rabi season, 2018 | | | Kharif season, 2019 | | |
|---|---|---|---|---|---|---|
| | Disease incidence in | | Disease severity (%) | Disease incidence in | | Disease severity (%) |
| | Plant (%) | Leaf (%) | | Plant (%) | Leaf (%) | |
| T$_1$ (Ridomil Gold) | 19.00 c | 20.00 c | 1.40 d | 15.00 c | 24.00 c | 1.76 cd |
| T$_2$ (Autostin 50WP) | 25.67 b | 29.00 b | 2.33 c | 17.67 bc | 29.00 bc | 2.76 bc |
| T$_3$ (Dithane M 45) | 29.00 ab | 24.33 bc | 3.00 b | 18.33 b | 28.33 c | 3.00 b |
| T$_4$ (Goldton 50WP) | 26.67 b | 24.33 bc | 2.80 bc | 20.00 b | 26.33 bc | 3.10 b |
| T$_5$ (Bordeaux Mixture) | 30.00 ab | 26.67 bc | 2.50 bc | 20.00 b | 31.00 bc | 3.00 b |
| T$_6$ (G-derma) | 29.33 ab | 26.00 bc | 2.50 bc | 20.00 b | 34.66 ab | 3.16 b |
| T$_7$ (Garlic bulb extract) | 29.00 ab | 25.00 bc | 2.67 bc | 20.00 b | 29.33 bc | 3.16 b |
| T$_8$ (Allamanda leaf extract) | 24.67 bc | 23.00 bc | 1.60 d | 14.67 c | 23.33 c | 1.50 d |
| T$_9$ (Control) | 35.00 a | 40.00 a | 4.00 a | 28.33 a | 41.66 a | 6.67 a |
| LSD (0.05) | 6.35 | 8.27 | 0.65 | 3.19 | 7.83 | 1.01 |
| CV (%) | 13.30 | 18.05 | 14.99 | 9.54 | 15.23 | 18.82 |

Means followed by the same letters in a column did not differ at 5% level of significance by LSD.

Table 5. Disease incidence and severity of white rust of red amaranth before 4th spray in the Rabi season of 2018 and the Kharif season of 2019.

| Treatments | Rabi season, 2018 | | | Kharif season, 2019 | | |
|---|---|---|---|---|---|---|
| | Disease incidence in | | Disease severity (%) | Disease incidence in | | Disease severity (%) |
| | Plant (%) | Leaf (%) | | Plant (%) | Leaf (%) | |
| $T_1$ (Ridomil Gold) | 26.00 d | 22.33 d | 1.76 d | 20.00 d | 27.33 c | 2.16 c |
| $T_2$ (Autostin 50WP) | 35.33 bc | 35.00 bc | 3.33 c | 29.00 c | 37.66 b | 3.66 b |
| $T_3$ (Dithane M 45) | 38.33 b | 32.33 bc | 4.33 b | 30.66 bc | 38.33 b | 4.00 b |
| $T_4$ (Goldton 50WP) | 37.33 bc | 32.67 bc | 4.16 bc | 33.33 bc | 35.66 b | 4.33 b |
| $T_5$ (Bordeaux Mixture) | 41.67 ab | 36.00 b | 3.50 bc | 32.33 bc | 38.33 b | 4.16 b |
| $T_6$ (G-derma) | 40.00 b | 32.67 bc | 3.60 bc | 34.00 b | 41.00 b | 4.33 b |
| $T_7$ (Garlic bulb extract) | 40.00 b | 32.67 bc | 4.00 bc | 34.00 b | 38.33 b | 4.00 b |
| $T_8$ (Allamanda leaf extract) | 31.33 cd | 27.67 cd | 2.16 d | 20.00 d | 25.66 c | 2.00 c |
| $T_9$ (Control) | 48.33 a | 51.00 a | 6.16 a | 46.66 a | 60.00 a | 10.33 a |
| LSD (0.05) | 6.79 | 7.40 | 0.92 | 4.49 | 5.39 | 1.06 |
| CV (%) | 10.44 | 12.73 | 14.55 | 8.35 | 8.20 | 14.26 |

Means followed by the same letters in a column did not differ at 5% level of significance by LSD.

Table 6. Disease incidence and severity of white rust of red amaranth after 7 days of 4[th] spray in the Rabi season of 2018 and the Kharif season of 2019.

| Treatments | Rabi season, 2018 | | | Kharif season, 2019 | | |
|---|---|---|---|---|---|---|
| | Disease incidence in | | Disease severity (%) | Disease incidence in | | Disease severity (%) |
| | Plant (%) | Leaf (%) | | Plant (%) | Leaf (%) | |
| $T_1$ (Ridomil Gold) | 24.00 c | 22.33 c | 1.83 d | 18.66 d | 24.00 c | 1.93 c |
| $T_2$ (Autostin 50WP) | 40.00 b | 36.67 b | 4.50 c | 31.66 c | 39.33 b | 4.33 b |
| $T_3$ (Dithane M 45) | 42.00 b | 38.33 b | 5.50 bc | 35.00 bc | 42.66 b | 4.83 b |
| $T_4$ (Goldton 50WP) | 42.00 b | 39.33 b | 5.50 bc | 33.66 bc | 42.33 b | 5.16 b |
| $T_5$ (Bordeaux Mixture) | 45.00 b | 41.00 b | 4.67 bc | 35.33 bc | 42.00 b | 4.83 b |
| $T_6$ (G-derma) | 42.00 b | 38.33 b | 4.67 bc | 37.33 b | 44.00 b | 5.06 b |
| $T_7$ (Garlic bulb extract) | 45.00 b | 40.00 b | 6.16 b | 37.66 b | 42.33 b | 4.83 b |
| $T_8$ (Allamanda leaf extract) | 31.00 b | 28.33 c | 2.43 d | 22.00 d | 27.66 c | 2.23 c |
| $T_9$ (Control) | 65.00 a | 60.00 a | 11.67 a | 61.66 a | 70.00 a | 16.66 a |
| LSD (0.05) | 8.11 | 8.16 | 1.60 | 4.60 | 5.52 | 1.06 |
| CV (%) | 11.18 | 12.33 | 17.81 | 7.65 | 7.67 | 11.09 |

Means followed by the same letters in a column did not differ at 5% level of significance by LSD.

Table 7. Reduction of incidence and severity of white rust of red amaranth after 7 days of 4$^{th}$ spray in the Rabi season of 2018 and the Kharif season of 2019.

| Treatments | Rabi season, 2018 | | | Kharif season, 2019 | | |
|---|---|---|---|---|---|---|
| | Reduction of disease incidence over control in | | Reduction of disease severity over control (%) | Reduction of disease incidence over control in | | Reduction of disease severity over control (%) |
| | Plant (%) | Leaf (%) | | Plant (%) | Leaf (%) | |
| T$_1$ (Ridomil Gold) | 63.07 | 62.78 | 84.31 | 69.73 | 65.71 | 88.41 |
| T$_2$ (Autostin 50WP) | 38.46 | 38.88 | 59.98 | 48.65 | 43.81 | 74.0 |
| T$_3$ (Dithane M 45) | 35.38 | 35 | 55.87 | 43.23 | 39.52 | 71.0 |
| T$_4$ (Goldton 50WP) | 35.38 | 36.11 | 56.98 | 45.41 | 39.67 | 69.02 |
| T$_5$ (Bordeaux Mixture) | 30.76 | 31.11 | 47.43 | 42.70 | 39.52 | 71.0 |
| T$_6$ (G-derma) | 35.38 | 36.11 | 56.98 | 39.45 | 37.14 | 69.20 |
| T$_7$ (Garlic bulb extract) | 30.76 | 33.33 | 51.21 | 38.92 | 39.52 | 71.0 |
| T$_8$ (Allamanda leaf extract) | 52.30 | 52.78 | 79.17 | 64.32 | 60.48 | 86.61 |

**Table 8. Effect of the different treatments on yield contributing characters and seed yield of red amaranth in the Rabi season of 2018**

| Treatments | Total no of plant/ plot | Plant height (cm) | No. of leaf/ plant | No. of inflorescence /plant | Seed yield/plot (g) | Seed yield/m$^2$ area (g) | Seed yield increase over control (%) |
|---|---|---|---|---|---|---|---|
| T$_1$ (Ridomil Gold) | 122.67 a | 64.66 a | 10.33 a | 8.33 a | 253 a | 84.33 | 53.32 |
| T$_2$ (Autostin 50WP) | 116.33 b | 58.66 bc | 7.67 bc | 7.0 b-d | 206 bc | 68.66 | 24.83 |
| T$_3$ (Dithane M 45) | 119.67 ab | 58.66 bc | 7.33 b-d | 7.33 a-c | 185 cd | 61.66 | 12.1 |
| T$_4$ (Goldton 50WP) | 119.0 ab | 58.66 bc | 7.33 b-d | 6.33 cd | 203 c | 67.66 | 23.01 |
| T$_5$ (Bordeaux Mixture) | 118.33 b | 60.33 b | 7.0 cd | 6.67 b-d | 186 cd | 62 | 12.72 |
| T$_6$ (G-derma) | 118.33 b | 55.67 c | 7.67 bc | 7.0 b-d | 186 cd | 62 | 12.72 |
| T$_7$ (Garlic bulb extract) | 118.00 b | 56.66 c | 6.67 d | 6.67 b-d | 183 cd | 61 | 10.9 |
| T$_8$ (Allamanda leaf extract) | 119.67 ab | 60.67 b | 8.0 b | 7.67 ab | 228 b | 76 | 38.18 |
| T$_9$ (Control) | 117.67 b | 56.67 c | 7.0 cd | 6.0 d | 165 d | 55 | - |
| LSD (0.05) | 4.29 | 3.09 | 0.91 | 1.09 | 24.88 | - | - |
| CV (%) | 2.09 | 3.03 | 6.87 | 9.07 | 7.19 | - | - |

Means followed by the same letters in a column did not differ at 5% level of significance by LSD.

**Table 9. Effect of the different treatments on yield contributing characters and seed yield of red amaranth in the Kharif season of 2019**

| Treatments | Total no of plant/ plot | Plant height (cm) | No. of leaf/ plant | No. of inflorescence /plant | Seed yield/plot (g) | Seed yield/$m^2$ area (g) | Seed yield increase over control (%) |
|---|---|---|---|---|---|---|---|
| $T_1$ (Ridomil Gold) | 116 a | 59.16 a | 9.33 a | 10.0 a | 232 a | 77.33 | 51.63 |
| $T_2$ (Autostin 50WP) | 112 bc | 55.16 b-d | 6.67 bc | 7.33 bc | 202 bc | 67.33 | 32.01 |
| $T_3$ (Dithane M 45) | 114 a-c | 55.50 bc | 6.33 bc | 7.33 bc | 180 cd | 60 | 17.64 |
| $T_4$ (Goldton 50WP) | 113 bc | 55.03 b-d | 6.33 bc | 6.33 cd | 175 cd | 58.33 | 14.37 |
| $T_5$ (Bordeaux Mixture) | 112 bc | 56.06 ab | 6.0 bc | 6.67 b-d | 173 cd | 57.66 | 13.05 |
| $T_6$ (G-derma) | 112 bc | 50.83 e | 6.67 bc | 7.0 b-d | 180 cd | 60 | 17.64 |
| $T_7$ (Garlic bulb extract) | 111 c | 52.70 c-e | 6.0 bc | 6.67 bd | 177 cd | 59 | 15.68 |
| $T_8$ (Allamanda leaf extract) | 114 ab | 56.67 ab | 7.33 b | 7.66 b | 210 b | 70 | 37.25 |
| $T_9$ (Control) | 113 bc | 52.0 de | 5.33 c | 6.0 d | 153 d | 51 | - |
| LSD (0.05) | 2.60 | 3.34 | 1.39 | 1.17 | 21.18 | - | - |
| CV (%) | 1.32 | 3.53 | 12.12 | 9.37 | 8.85 | - | - |

Means followed by the same letters in a column did not differ at 5% level of significance by LSD.